\documentclass[showpacs,amsmath,twocolumn,amssymb]{revtex4}
\usepackage{graphicx}
\usepackage{dcolumn}
\usepackage{bm}

\begin{document}

\title{Convergence of Perturbations for a Big Bounce in Loop Quantum Cosmology}
\author{Yu Li}
\email{leeyu@mail.bnu.edu.cn}
  \affiliation{Department of Physics, Beijing Normal University, Beijing 100875, China}
\author{Jian-Yang Zhu}
\thanks{Author to whom correspondence should be addressed}
\email{zhujy@bnu.edu.cn}
  \affiliation{Department of Physics, Beijing Normal University, Beijing 100875, China}
\date{\today}

\begin{abstract}
We investigate the convergence behaviors of the scalar and the
vector perturbations for a big bounce phase in loop quantum
cosmology. Two models are discussed: one is the universe filled by a
massless scalar field; the other is a toy model which is
radiation-dominated in the asymptotic past and future. We find that
the behaviors of the Bardeen potential of the scalar mode near both
the bounce point and the transition point of the null energy
condition are good, moreover, the unlimited growth of the vector
perturbation can be avoided in our bounce model. This is different
from the bounce models in pure general relativity. And we also find
that the maximum of an observable vector mode is inversely
proportional to the square of the minimum scalar factor
$a_{bounce}$. This conclusion is independent with the bounce model,
and we may conclude that the bounce in loop quantum cosmology is
reasonable.

\end{abstract}

\maketitle


\section{\label{s1}Introduction}
There are several ways to solve the singularity problem in cosmology
\cite{r7}, one of them is the bounce model \cite{r8,r9,r10}. In the
studying of the bounce model in pure general relativity (GR)
\cite{r11}, the scalar hydrodynamic perturbation can lead to a
singular behavior of the Bardeen potential. The further study
\cite{r12,r13,r14} show that the divergence of the Bardeen potential
can be avoided only in some models with special matter.

Another mode of perturbations must be considered is the vector
perturbation. It is known that the vector mode will decay quickly in
expanding phase of the universe. So it exhibits a growing mode
solution in a contracting universe \cite{r97}. This growing will in
general lead to the breakdown of perturbation theory near the
bounce. So, it is necessary to check the behaviors of the vector
perturbations near the bounce.

Generally speaking, the bouncing phase originates from a quantum
effect of gravity. So it is interesting to study the bounce model in
quantum gravity theory.

At present, the issue of finding a complete theory of quantum
gravity is still open. In current approaches, one of the most active
is loop quantum gravity. Loop quantum gravity (LQG)
\cite{lqg1,lqg2,lqg3} is a mathematically well-defined,
non-perturbative and background independent quantization of general
relativity. And, Loop Quantum Cosmology (LQC) \cite{B}, a symmetry
reduction of LQG to the homogeneous and isotropic spacetime, has
achieved many successes. A major success of LQC is the resolution of
the Big Bang singularity \cite{bb,nbb1,nbb2}, this result depends
crucially on the discreteness of the spacetime. Instead of Big Bang,
there will be a Big Bounce.

In GR bounce, the divergence of the Bardeen potential occurs at two
point \cite{r11}, i.e., near the bounce point and near the
transition point of null energy condition (NEC). The difference is
that the LQC bounce is governed by a discrete quantum geometry
\cite{r15}. This will lead to some different behaviors of the
Bardeen potential. In this paper, we consider the behaviors of the
perturbations near both the bounce point and the transition point of
NEC under the framework of the effective theory of LQC.

The paper is organized as follows. In Sec. \ref{s2}, we give the
framework of the effective theory of LQC with holonomy corrections,
it can yield the bounce background. In Sec.\ref{s3}, we introduce
the scalar perturbation based on the Sec. \ref{s2}. Two models are
analyzed in this section: one is the universe filled by a massless
scalar field; the other is a toy model which is radiation-dominated
in the asymptotic past and future. In Sec. \ref{s6} we discuss the
vector perturbation near the bounce. The discussion and conclusions
are presented in Sec. \ref{s7}.

\section{\label{s2}background of LQC bounce}
The canonical variables used in LQG are the Ashtekar connection
$A_a^i$ and the densitized triad $E_i^a$ \cite{lqg1,lqg2,lqg3},
where $A_a^i=\Gamma_a^i+K_a^i$ with $\Gamma$ the spin connection and
$K$ the extrinsic curvature, and $E_i^a=e_i^a/|dete_i^b|$ with
$e_i^ae_i^b=q^{ab}$ and $q_{ab}$ the spatial metric. For a spatially
homogeneous and isotropic universe model (FRW metric), the Ashtekar
variables can be reduced to the diagonal form, i.e.,
$A^i_a=c\delta^i_a$ and $E^a_i=p\delta^a_i$ \cite{B}. Therefore, the
basic canonical variables for the gravitational field are
$(\bar{c},\bar{p})$ and for the scalar field
$(\bar{\varphi},p_{\bar{\varphi}})$. Here we denote the background
variables with a bar. In this paper, we consider only the flat space
universe. Thus the canonical variables can be expressed in terms of
the standard FRW variables as:
$(\bar{c},|\bar{p}|)=(\gamma\dot{a},a^2)$, where $a$ is the scale
factor; and the effective Hamlitonian of the considered model is
given by
\begin{equation}\label{e1}
H=-\frac{3}{8\pi G \gamma^2}\sqrt{|\bar{p}|}\bar{c}^2+\frac{1}{2}
\frac{p_{\bar{\varphi}}^2}{|\bar{p}|^{3/2}}+|\bar{p}|^{3/2}V(\bar{\varphi}),
\end{equation}
where the factor $\gamma$ is called the Barbero-Immirzi parameter
which is a constant of the theory.

In the process of quantization, we can find that there is no
operator corresponding to the canonical variable $\bar{c}$ itself
but we can return to the holonomy. This fact can lead to the
so-called holonomy correction in an effective theory of LQC. The
effects of this correction can be obtain by simply replacing the
$\bar{c}$ to $\sin (\tilde{\mu}\bar{c})/\tilde{\mu}$ with the choice
of\footnote{In  \cite{Li},  the lattice power law of Loop Quantum
Cosmology, $\tilde{\mu}\propto p^{\beta}$, has been analysed by
applying the higher order holonomy correction to the perturbation
theory of cosmology, and the range of $\beta$ has been decided to be
[-1,0].}
\begin{equation}\label{e2}
\tilde{\mu}=\sqrt{\frac{\Delta}{|\bar{p}|}},
\end{equation}
where $\Delta\equiv 2\sqrt{3}\pi\gamma l_p^2$ is a area gap, and
$l_p$ denotes the Planck length. So, the effective Hamiltonian with
holonomy correction is
\begin{equation}\label{e3}
H_{eff}=-\frac{3}{8\pi G \gamma^2}\sqrt{|\bar{p}|}\left[\frac{\sin (\tilde{\mu}\bar{c})}{\tilde{\mu}}\right]^2+\frac{1}{2}\frac{p_{\bar{\varphi}}^2}{|\bar{p}|^{3/2}}+|\bar{p}|^{3/2}V(\bar{\varphi}).
\end{equation}
From now on, we focus on a positive $\bar{p}$.

The equations of motion can now be derived by the using of the
Hamilton equation
\begin{equation}\label{e4}
\dot{f}=\{f,H_{eff}\},
\end{equation}
where the dot denotes the derivative with respect to the cosmic time
$t$ and the Poisson bracket is defined as
\begin{eqnarray}\label{e5}
\{f,g\}=&&\frac{8\pi G \gamma}{3}\left[\frac{\partial f}{\partial \bar{c}}\frac{\partial g}{\partial \bar{p}}-\frac{\partial f}{\partial \bar{p}}\frac{\partial g}{\partial \bar{c}}\right]\nonumber\\
&+&\left[\frac{\partial f}{\partial \bar{\varphi}}\frac{\partial g}{\partial p_{\bar{\varphi}}}-\frac{\partial f}{\partial p_{\bar{\varphi}}}\frac{\partial g}{\partial \bar{\varphi}}\right].
\end{eqnarray}

From this definition we can obtain two elementary brackets
\begin{equation}\label{e6}
\{\bar{c},\bar{p}\}=\frac{8\pi G \gamma}{3} ,\ \  \ \ \{\bar{\varphi},p_{\bar{\varphi}}\}=1.
\end{equation}
By using these brackets, one can derive two equations, i.e., an
effective Friedmann equation and an effective Raychaudhuri equation.
However, in theory of perturbation, the use of the conformal time
$\eta$ may be convenient than the cosmic time. The conformal time
can be related to the cosmic time $t$ through the scale factor $a$,
$ad\eta=dt$. Thence the effective Friedmann equation and the
effective Raychaudhuri equation with conformal time are respectively
\cite{r1},
\begin{equation}\label{e7}
\left[\frac{\sin (\tilde{\mu}\gamma\bar{\mathfrak{K}})}{\gamma\tilde{\mu}}\right]^2=\frac{8\pi G }{3}\left[\frac{1}{2}(\bar{\varphi}')^2+\bar{p}V(\bar{\varphi})\right],
\end{equation}
\begin{eqnarray}\label{e8}
\bar{\mathfrak{K}}'&+&\frac{1}{2}\left[\frac{\sin (\tilde{\mu}\gamma\bar{\mathfrak{K}})}{\gamma\tilde{\mu}}\right]^2+\bar{p}\frac{\partial}{\partial \bar{p}}\left[\frac{\sin (\tilde{\mu}\gamma\bar{\mathfrak{K}})}{\gamma\tilde{\mu}}\right]^2\nonumber\\
&&=4\pi
G\left[-\frac{1}{2}(\bar{\varphi}')^2+\bar{p}V(\bar{\varphi})\right],
\end{eqnarray}
where the prime denotes the derivative with respect to the conformal
time $\eta$ and $\gamma\bar{\mathfrak{K}}\equiv \bar{c}$. In
addition, the Klein-Gordon equation can also be derived as follows:
\begin{equation}\label{e9}
\bar{\varphi}''+2\left[\frac{\sin (2\tilde{\mu}\gamma\bar{\mathfrak{K}})}{2\gamma\tilde{\mu}}\right]\bar{\varphi}'+\bar{p}V_{,\bar{\varphi}}(\bar{\varphi})=0.
\end{equation}

From Eq.(\ref{e4}) and the relation between cosmic time and conformal time, one can get the motion equation of $\bar{p}$ with conformal time
\begin{equation}\label{e10}
\bar{p}'=2\bar{p}\left[\frac{\sin (2\tilde{\mu}\gamma\bar{\mathfrak{K}})}{2\gamma\tilde{\mu}}\right].
\end{equation}
So, one can further define a conformal Hubble parameter $\mathbb{H}$
by
\begin{equation}\label{e11}
\mathbb{H}:=\frac{\bar{p}'}{2\bar{p}}=\frac{\sin (2\tilde{\mu}\gamma\bar{\mathfrak{K}})}{2\gamma\tilde{\mu}}.
\end{equation}
Moreover, we can also define
\begin{eqnarray}
\mathfrak{S}_1&:= & \mathbb{H}^2-\left[\frac{\sin (\tilde{\mu}\gamma\bar{\mathfrak{K}})}{\gamma\tilde{\mu}}\right]^2,\label{e12}\\
\mathfrak{S}_2&:=
&\mathbb{H}'-\bar{\mathfrak{K}}'-\bar{p}\frac{\partial}{\partial
\bar{p}}\left[\frac{\sin
(\tilde{\mu}\gamma\bar{\mathfrak{K}})}{\gamma\tilde{\mu}}\right]^2.\label{e22}
\end{eqnarray}
In fact, $\mathfrak{S}_1$ and $\mathfrak{S}_2$ can be taken as two
effective corrections to the evolution equations of the Bardeen
potential. This is because that one can obtain the classical
evolution equations by taking $\mathfrak{S}_1\rightarrow0$ and
$\mathfrak{S}_2\rightarrow0$ at the same time.

We rewrite the Eqs.(\ref{e7}) and (\ref{e8}) in terms of
$\mathfrak{S}_1$, $\mathfrak{S}_2$ and $\mathbb{H}$
\begin{eqnarray}
\mathbb{H}^2-\mathfrak{S}_1=\frac{8\pi G}{3}\bar{p}\times\rho_s,\label{e13}\\
-\frac{1}{3}(\mathbb{H}^2-\mathfrak{S}_1)-\frac{2}{3}(\mathbb{H}'-\mathfrak{S}_2)
=\frac{8\pi G}{3}\bar{p}\times P_s,\label{e15}
\end{eqnarray}
where
\begin{equation}
\rho_s:=\frac{(\bar{\varphi}')^2}{2\bar{p}}+V(\bar{\varphi}),\label{e14}
\end{equation}
and
\begin{equation}
 P_s:=\frac{(\bar{\varphi}')^2}{2\bar{p}}-V(\bar{\varphi})
\end{equation}
are the energy density and pressure of scalar field respectively.

By using Eqs.(\ref{e11}) and (\ref{e14}), one can rewrite the
Eq.(\ref{e13}) to
\begin{equation}\label{e61}
\mathbb{H}^2=l_p^2 \bar{p}\
\rho_s\left(1-\frac{\rho_s}{\rho_c}\right),
\end{equation}
where
\begin{equation}\label{e62}
\rho_c=\frac{1}{l_p^2\gamma^2\Delta}.
\end{equation}

Obviously, one can easily check that when
$\rho_s=\rho_c$, $\mathbb{H}=0$, and this means that a
bounce occurs. The bounce density $\rho_c$ is related to
$\Delta$ which is the smallest eigenvalue of area operator, so this
LQC bounce is originated from the quantum effects of spacetime.

From Eqs.(\ref{e13}) and (\ref{e15}) we can get a useful relation
equation
\begin{equation}\label{e16}
\rho_s+ P_s=\frac{1}{4\pi G \bar{p}}\beta,
\end{equation}
where
\begin{equation}\label{e17}
\beta:=\mathbb{H}^2-\mathbb{H}'-\mathfrak{S}_1+\mathfrak{S}_2.
\end{equation}
Thus, the relationship \cite{r3} between the null energy condition
and the stress-energy tensor can be written as:
\begin{equation}\label{e18}
NEC\Longleftrightarrow\rho_s+ P_s\geqslant0\Longleftrightarrow\beta\geqslant0.
\end{equation}

\section{\label{s3}scalar perturbation with holonomy corrections}
In this section, we introduce the scalar perturbation based on the
Sec. \ref{s2} and analyze in detail the following two models: one is
the universe filled by a massless scalar field; the other is a toy
model which is radiation-dominated in the asymptotic past and
future. In our models, the spacetime is described by the metric
\begin{equation}\label{e19}
ds^2=a^2(\eta)\left[-(1+2\Phi)d\eta^2+(1-2\Psi)\delta_{ab}dx^adx^b\right],
\end{equation}
where $\Phi$ and $\Psi$ are the Bardeen potential. In the case of
vanishing anisotropic stresses, $\Phi$ and $\Psi$ are equal
\cite{r4}. Therefore, we set $\Phi=\Psi$ from now on. The evolution
equations of the Bardeen potential with holonomy corrections have
been given in \cite{r1}:
\begin{widetext}
\begin{eqnarray}
\nabla^2\Phi-3\mathbb{H}\Phi'-\left[\bar{\mathfrak{K}}'+6\mathbb{H}^2-4\left[\frac{\sin (\tilde{\mu}\gamma\bar{\mathfrak{K}})}{\gamma\tilde{\mu}}\right]^2
+\bar{p}\frac{\partial}{\partial \bar{p}}\left[\frac{\sin (\tilde{\mu}\gamma\bar{\mathfrak{K}})}{\gamma\tilde{\mu}}\right]^2\right]\Phi&=&4\pi G\bar{p}\ \delta\rho_s, \nonumber\\
\label{e20}\\
\Phi''+\left\{3\mathbb{H}+\frac{1}{\mathbb{H}}\left[\mathbb{H}'-\bar{\mathfrak{K}}'-\bar{p}\frac{\partial}{\partial
\bar{p}}\left[\frac{\sin
(\tilde{\mu}\gamma\bar{\mathfrak{K}})}{\gamma\tilde{\mu}}\right]^2\right]\right\}\Phi'+\left[2\mathbb{H}^2+4\mathbb{H}'-3\bar{\mathfrak{K}}'
-3\bar{p}\frac{\partial}{\partial \bar{p}}\left[\frac{\sin
(\tilde{\mu}\gamma\bar{\mathfrak{K}})}{\gamma\tilde{\mu}}\right]^2\right]\Phi&=&4\pi
G\bar{p}\ \delta P_s. \nonumber\\
 \label{e21}
\end{eqnarray}
\end{widetext}

By Using the definitions of Eqs.(\ref{e12}) and (\ref{e22}), we can
rewrite the Eqs.(\ref{e20}) and (\ref{e21}) as
\begin{eqnarray}
\nabla^2\Phi-3\mathbb{H}\Phi'-\left[2\mathbb{H}^2+\mathbb{H}'-4\mathfrak{S}_1
-\mathfrak{S}_2\right]\Phi&=&4\pi G\bar{p}\ \delta\rho_s,\nonumber\\
\label{e23}\\
\Phi''+\left(3\mathbb{H}+\frac{\mathfrak{S}_2}{\mathbb{H}}\right)\Phi'
+\left[2\mathbb{H}^2+\mathbb{H}'+3\mathfrak{S}_2\right]\Phi&=&4\pi G\bar{p}\ \delta P_s.\nonumber\\
\label{e24}
\end{eqnarray}

We want to get the evolution equation of the Bardeen potential but
the matter is not contained in it. So we should obtain the
relationship between $\delta\rho_s$ and $\delta P_s$.
In general, the pressure perturbation can be separated into two
parts of the adiabatic and the entropic perturbation as follows
\begin{equation}\label{e25}
\delta P_s=\left(\frac{\partial  P_s}{\partial
\rho_s}\right)_S \delta\rho_s+\left(\frac{\partial
 P_s}{\partial S}\right)_{\rho_s} \delta
S=\Upsilon\delta\rho_s+\tau\delta S.
\end{equation}
For the hydrodynamic matter, $\Upsilon$ can be interpreted as the
sound velocity. In this paper, we focus on a adiabatic perturbation
only, so we have
\begin{equation}\label{e26}
\Upsilon=\frac{\delta P_s}{\delta\rho_s}=\frac{ P_s'}{\rho_s'}.
\end{equation}
From Eqs.(\ref{e13}), (\ref{e15}) and (\ref{e17}) we can obtain
\begin{equation}\label{e27}
\Upsilon=-\frac{1}{3}\left(1+\frac{2\beta'+\mathfrak{S}_1'-2\mathbb{H}\mathfrak{S}_2}
{2\mathbb{H}\beta+\mathfrak{S}_1'-2\mathbb{H}\mathfrak{S}_2}\right).
\end{equation}
Using Eq.(\ref{e26}) and inserting Eq.(\ref{e24}) into
Eq.(\ref{e23}), one gets
\begin{eqnarray}
0 &=&\Phi ^{\prime \prime }+\left[ 3\mathbb{H}(1+\Upsilon)+\frac{\mathfrak{S}_2}{%
\mathbb{H}}\right] \Phi ^{\prime }-\Upsilon \nabla ^2\Phi  \nonumber \\
&&+\left[(1+\Upsilon )\left( 2\mathbb{H}^2+\mathbb{H}^{\prime }\right) -4\Upsilon %
\mathfrak{S}_1+\left( 3-\Upsilon \right) \mathfrak{S}_2\right] \Phi
.  \nonumber \\
\label{e28}
\end{eqnarray}
The equation for the mode of wave number $k$ is
\begin{eqnarray}
0 &=&\Phi _k^{\prime \prime }+\left[ 3\mathbb{H}(1+\Upsilon)+\frac{%
\mathfrak{S}_2}{\mathbb{H}}\right] \Phi _k^{\prime }+\left[ \Upsilon
k^2\right.   \nonumber \\
&&\left. +\left( 1+\Upsilon \right) \left(
2\mathbb{H}^2+\mathbb{H}^{\prime
}\right) -4\Upsilon \mathfrak{S}_1+\left( 3-\Upsilon \right) \mathfrak{S}%
_2\right] \Phi _k.  \nonumber \\
\label{e29}
\end{eqnarray}
\subsection{\label{s4}Massless scalar field}
In this subsection, we discuss a simple model which the universe is
filled with a massless scalar field, i.e., $V(\varphi)=0$. The exact
solution of $\bar{p}(t)$ is \cite{r6}
\begin{equation}\label{e30}
\bar{p}(t)=(\bar{p}_{min}^3+\mathcal {A}t^2)^{1/3},
\end{equation}
where
\begin{equation}\label{e31}
\bar{p}_{min}=\gamma
l_p^2\left[\frac{8\sqrt{3}}{3}\pi^2\left(\frac{p_{\bar{\varphi}}}{l_p}\right)^2\right]^{1/3}
\end{equation}
is the minimal value of $\bar{p}(t)$, and
\begin{equation}\label{e32}
\mathcal{A}=12\pi G p_{\bar{\varphi}}^2.
\end{equation}
Because $V(\varphi)=0$, one can get $\dot{p}_{\bar{\varphi}}=0$,
$p_{\bar{\varphi}}=const.$, then $\bar{p}_{min}$ and $\mathcal{A}$
are constant too.

In our discussion, we need to know the evolution function of
$\bar{p}$ with $\eta$. It can be obtained by using the relation of
$ad\eta=dt$. So, we have
\begin{equation}\label{e33}
\eta=\int\frac{1}{\bar{p}(t)^{1/2}}dt.
\end{equation}

However, the result of the integration is complex. To keep our
discussion simple, we consider a asymptotic behavior of
$\bar{p}(t)$:
\begin{equation}\label{e34}
\bar{p}(t)=\left\{
\begin{aligned}
\mathcal{A}^{1/3}t^{2/3}, \ \ \ |t|\rightarrow\infty,\ \ \ (far\  from\ the\ bounce)\\
\bar{p}_{min}, \ \ |t|\rightarrow0.\ \ \ \ \ \ \ \ \ \ \ (near\ the\
bounce)
\end{aligned}
\right.
\end{equation}

From the integration of Eq.(\ref{e33}), we can get the asymptotic
behavior of $\eta$:
\begin{equation}\label{e35}
|\eta|=\left\{
\begin{aligned}
\frac{3}{2}\mathcal{A}^{-1/6}t^{2/3}, \ \ \ |t|\rightarrow\infty,\ \ \ (far\  from\ the\ bounce)\\
\bar{p}_{min}^{-1/2}t, \ \ |t|\rightarrow0.\ \ \ \ \ \ \ \ \ \ \
(near\ the\ bounce)
\end{aligned}
\right.
\end{equation}
Then, we have an approximate relation between $\eta$ and $t$
\begin{equation}\label{e36}
|\eta|\approx \frac{3}{2}\mathcal{A}^{-1/6}t^{2/3}.
\end{equation}

\begin{figure}[h!]
\includegraphics[width=0.48\textwidth]{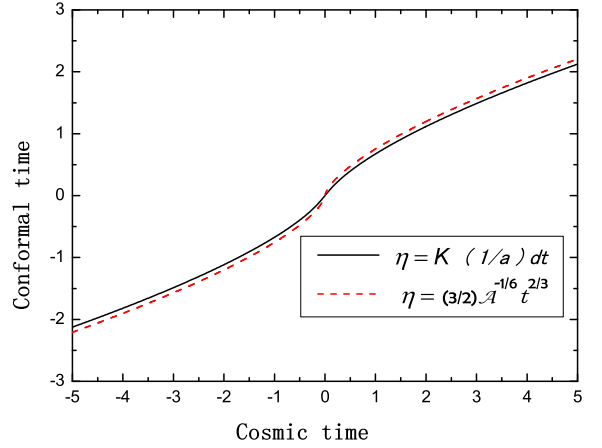}
\caption{The relation between the conformal time $\eta$ and the
cosmic time $t$. The solid line is $\eta=\int (1/a) dt$, and the
dash line is $\eta=(3/2)\mathcal{A}^{-1/6}t^{2/3}$.}\label{f1}
\end{figure}
Fig.\ref{f1} show that $|\eta|=(3/2)\mathcal{A}^{-1/6}t^{2/3}$ is a
good approximation for $\eta=\int (1/a) dt$. The most
straightforward way is to insert this approximate relation to
Eq.(\ref{e30}). However, this relation is only a asymptotic behavior
corresponding to the case of $|t|\rightarrow\infty$, not the
behavior on all time. So we should consider the asymptotic behavior
of $\bar{p}(\eta)$.

Under the approximation of Eq.(\ref{e36}), the asymptotic behavior
of $\bar{p}(\eta)$ is
\begin{equation}\label{e37}
\bar{p}(\eta)=\left\{
\begin{aligned}
\frac{2}{3}\mathcal{A}^{1/2}|\eta|, \ \ \ |t|\rightarrow\infty,\ \ \ (far\  from\ the\ bounce)\\
\bar{p}_{min}, \ \ |t|\rightarrow0.\ \ \ \ \ \ \ \ \ \ \ (near\ the\
bounce)
\end{aligned}
\right.
\end{equation}

Now, we construct a function of $\bar{p}(\eta)$ approximately. The
approximate function should satisfy the asymptotic behavior of
$\bar{p}(\eta)$. One class of such functions are
\begin{equation}\label{e38}
\bar{p}(\eta)=\left[\bar{p}_{min}^n+\left(\frac{2}{3}\right)^n\mathcal{A}^{n/2}|\eta|^n\right]^{1/n},
\end{equation}
where $n$ is a natural number.

One can find that if $n$ is even, there will not be an absolute
value like $|\eta|$ appeared in the equation. So, from now on we
set $n=2$. There is another reason to choose $n=2$ that only $n=2$
can lead to a evolution equation of the Bardeen potential which have
analytical solution (see Eq.(\ref{e44})).

So we set the approximate function of $\bar{p}(\eta)$ is
\begin{equation}\label{e39}
\bar{p}(\eta)=\left(\bar{p}_{min}^2+\mathbb{A}\eta^2\right)^{1/2},
\end{equation}
where $\mathbb{A}=\frac{4}{9}\mathcal{A}$. Eq.(\ref{e39}) is
different with Eq.(\ref{e30}). Equation (\ref{e30}) is exact
evolution of $\bar{p}$ but Eq.(\ref{e39}) is a approximation of
Eq.(\ref{e30}).

Under the condition of $V(\bar{\varphi})=0$, we have
$ P_s=\rho_s$, so $\Upsilon=1$. From Eqs.(\ref{e13}),
(\ref{e15}) and (\ref{e17}) we can obtain
\begin{eqnarray}
\beta&=&3(\mathbb{H}^2-\mathfrak{S}_1), \label{e40}\\
\mathfrak{S}_2&=&2\mathbb{H}^2-2\mathfrak{S}_1+\mathbb{H}'.\label{e41}
\end{eqnarray}
Using these equations and Eq.(\ref{e29}) one can get
\begin{equation}\label{e42}
\Phi''_k+\left(8\mathbb{H}+\frac{\mathbb{H}'}{\mathbb{H}}
-2\frac{\mathfrak{S}_1}{\mathbb{H}}\right)\Phi'_k+(k^2+4\mathfrak{S}_1)\Phi_k=0.
\end{equation}
Using Eq.(\ref{e39}) and Eq.(\ref{e42}), we can discuss the
behaviors of the evolution equation of the Bardeen potential in the
following cases.
\subsubsection{Near the bounce}
Near the bounce point, $|\eta|\rightarrow 0$, and the leading order
of the coefficients of evolution equation are
\begin{equation}
8\mathbb{H}+\frac{\mathbb{H}^{\prime }}{\mathbb{H}}-2\frac{\mathfrak{S}_1}{%
\mathbb{H}}\approx \left(1+\frac{2p_{\bar{\varphi}}^2l_p^2}{\mathbb{A}}\right)\frac 1\eta
,\quad 4\mathfrak{S}_1\approx -\frac{2p_{\bar{\varphi}}^2}{l_p^2}.
\label{e43}
\end{equation}
And the evolution equation changes to
\begin{equation}\label{e44}
\Phi''_k+\left(1+\frac{2p_{\bar{\varphi}}^2
l_p^2}{\mathbb{A}}\right)
\frac{1}{\eta}\Phi'_k+\left(k^2-\frac{2p_{\bar{\varphi}}^2}{l_p^2}\right)\Phi_k=0.
\end{equation}
and the solutions of Eq.(\ref{e44}) is
\begin{equation}\label{e86}
\Phi_k(\eta)=\eta^{\nu}[C_1Z_{\nu}(\mathcal{K}\eta)+C_2Z_{-\nu}(\mathcal{K}\eta)]
\end{equation}

where
\begin{equation}
\mathcal{K}=\sqrt{k^2-\frac{2p_{\varphi}^2}{l_p^2}},\ \ \ \nu=-\frac{p_{\bar{\varphi}}^2 l_p^2}{\mathbb{A}}.
\end{equation}
and $Z_{\nu}$ is Bessel function, $C_1$, $C_2$ are arbitrary constants.

The leading order of this solution is
\begin{equation}\label{e45}
\Phi_k(\eta)\approx\Phi_{(1)}+\Phi_{(2)}\eta^{2\nu},
\end{equation}
where $\Phi_{(1)}$ and $\Phi_{(2)}$ are constants which related to $k$.

One can found that, $\nu<0$, and the second term in Eq.(\ref{e45})
is  divergence. But we can choose the arbitrary constant $C_2=0$,
which means $\Phi_{(2)}=0$. Thus we can obtain a convergence
solution of scalar perturbations near the bounce.

\subsubsection{NEC transition problem}
In our model, the point of NEC transition is obtained form $\beta=0$. When $V(\bar{\varphi})=0$, the $\beta$ is Eq.(\ref{e40}).
\begin{figure}[h!]
\includegraphics[width=0.48\textwidth]{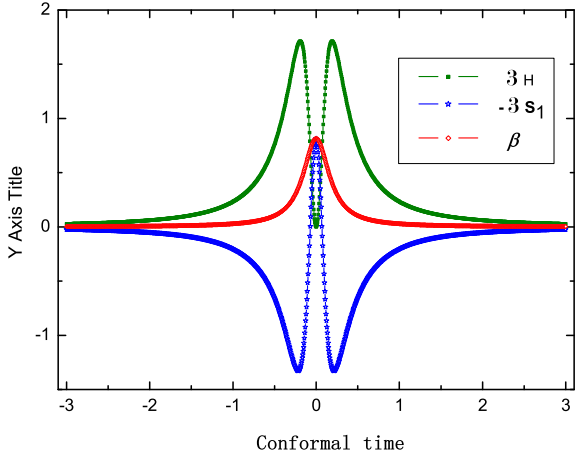}
\caption{The evolutions of $\mathbb{H}$, $-\mathfrak{S}_1$ and
$\beta$ near the bounce point.}\label{f2}
\end{figure}
From the Fig.\ref{f2}, we can see that $\beta$ is always positive in
our model. So there is no NEC transition. And then, there is not the
problem of the divergence near the point of NEC transition.

In GR, if the matter always satisfies NEC, there will be no bounce.
The reason is that the GR bounce is led by some exotic matter which
violates the NEC. However, in LQC, the bounce is originated in the
discrete spacetime geometry. It is a quantum effect of spacetime.
So, even if the matter never violates the NEC, there is also a
bounce phase.

\subsection{\label{s5}Toy model}
In this subsection, we extend our discussion slightly, and consider
a toy model which was introduced in \cite{r98,r99}. In this model,
the universe is radiation-dominated in the asymptotic past and
future. In other words, the asymptotic behavior of $a(\eta)$ should
be $a\propto\eta$ or $\bar{p}\propto\eta^2$. So we assume that there
are some $V(\bar{\varphi})$ that can make the form of
$\bar{p}(\eta)$ as\footnote{In \cite{r99,r98} the evolution of scale
factor is $a(\eta)=\sqrt{b+\alpha\eta^2}$, we generalize it a
little.  }
\begin{equation}\label{e49}
\bar{p}(\eta)=(b^m+\alpha\eta^{2m})^{1/m},
\end{equation}
where $b$, $\alpha$ are constants, and $b>0,m>1$.

\subsubsection{Near the bounce}
Under the assumption of Eq.(\ref{e49}), we can obtain the evolution
equation of the perturbation near the bounce point from
Eq.(\ref{e29})
\begin{eqnarray}\label{e50}
\Phi_k''&+&\left[1-\frac{2m-2}{\sqrt{b}}-(2m-1)(\sqrt{b}+b)\right]\frac{1}{\eta}\Phi_k'\nonumber\\
&-&\frac{b^{(2m-1)/2}(2m-2)(1+\sqrt{b})}{3\alpha}\frac{k^2}{\eta^{2m}}\Phi_k=0.
\end{eqnarray}
In fact, Eq.(\ref{e50}) is only accurate to its leading order. We
can obtain the following analytical solution
\begin{equation}\label{e51}
\Phi_k(\eta)=\eta^{(1-m)\nu}[\Phi_{(1)}\mathcal{Z}_{\nu}(\kappa\eta^{1-m})
+\Phi_{(2)}\mathcal{Z}_{-\nu}(\kappa\eta^{1-m})],
\end{equation}
where $\mathcal {Z}_{\nu}$ is the Bessel function, $\Phi_{(1)}$ and
$\Phi_{(2)}$ are constants, and
\begin{eqnarray}
\nu&=&\frac{1}{1-m}\left[\frac{m-1}{\sqrt{b}}+\left(n-\frac{1}{2}\right)(\sqrt{b}+b)\right],\\
\kappa&=&\frac{ik}{m-1}\sqrt{\frac{b^{(2m-1)/2}(2m-2)(1+\sqrt{b})}{3\alpha}}.
\end{eqnarray}
The leading order of Eq.(\ref{e51}) is
\begin{equation}\label{e52}
\Phi_k(\eta)\approx\kappa^{\nu}[\Phi_{(1)}+\Phi_{(2)}\eta^{2(1-m)\nu}].
\end{equation}
Because $2(1-m)\nu=(2m-2)/\sqrt{b}+(m-1)(\sqrt{b}+b)>0$, so the
behavior of the perturbation near the bounce point is good.

\subsubsection{NEC transition problem}
Near the bounce point, the leading order of $\beta$ is
\begin{equation}\label{e53}
\beta\approx-\frac{\alpha(2m-1)(\sqrt{b}+b)}{b^m}\eta^{2m-2}<0.
\end{equation}
So, there must be a time point corresponding to $\beta=0$. To get
this point, we consider the next order of $\beta$
\begin{eqnarray}\label{e54}
\beta\approx&-&\frac{\alpha(2m-1)(\sqrt{b}+b)}{b^m}\eta^{2m-2}\nonumber\\
&+&\frac{\alpha^2[(2m-1)(4m-1)(\sqrt{b}+b)+b]}{2mb^{2m}}\eta^{4m-2}.\nonumber\\
\end{eqnarray}
Near the bounce point, the leading order term is $\eta^{2m-2}$ term,
so $\beta<0$. When $|\eta|$ get larger, the leading order term will
be $\eta^{4m-2}$, so $\beta>0$. The transition point is $\beta=0$,
it leads to a transition point
\begin{equation}\label{e55}
\eta_0\approx\pm\left[\frac{b^m}{\alpha(4m-1)}\right]^{\frac{1}{2m}},
\end{equation}
$\eta_0$ denotes the transition time.

When we discuss the perturbation near the $\eta_0$, we can shift the
origin of the time such that $\eta_0=0$. If we do that, the function
of $\bar{p}(\eta)$ will be
\begin{equation}\label{e56}
\bar{p}(\eta)=[b^m+\alpha(\eta+\eta_0)^{2m}]^{1/m}.
\end{equation}
Near the point $\eta_0$, $\beta\approx0$ and $\Upsilon\approx-2/3$.
Moreover, from the definition of $\beta$ and $\beta=0$, we have
$\mathfrak{S}_2=\mathfrak{S}_1-\mathbb{H}^2+\mathbb{H}'$. Thus,
Eq.(\ref{e29}) changes to
\begin{equation}\label{e57}
\Phi_k''+\frac{\mathfrak{S}_1+\mathbb{H}'}{\mathbb{H}}\Phi_k'+\left(-\frac{2}{3}k^2
+5\mathfrak{S}_1-\frac{5}{3}\mathbb{H}^2+\frac{8}{3}\mathbb{H}'\right)\Phi_k=0.
\end{equation}
Near the $\eta_0$, the leading order of this equation is
\begin{equation}\label{e58}
\Phi_k''+L_1\Phi_k'+\left(-\frac{2}{3}k^2+L_2\right)\Phi_k=0,
\end{equation}
where $L_1$ and $L_2$ are constants which can be related to
$m,\Delta$ etc. The solution of this equation is
\begin{equation}\label{e59}
\Phi_k=\Phi_{(1)}\exp\left[-\frac{1}{2}(L_1+\hat{k})\eta\right]+\Phi_{(2)}\exp\left[-\frac{1}{2}(L_1-\hat{k})\eta\right],
\end{equation}
where $\Phi_{(1)}$ and $\Phi_{(2)}$ are constants again and
\begin{equation}\label{e60}
\hat{k}=\sqrt{4\left(\frac{2}{3}k^2-L_2\right)+L_1^2}.
\end{equation}
Near the NEC transition point, we can see a convergence behavior of
the perturbation.

\section{\label{s6}vector perturbation with holonomy corrections}
In the most of research on perturbations of cosmology, a lot of
attention have been focused on the scalar and the tensor mode. The
reason is that the vector mode will decay quickly in expanding phase
of universe.

However, in the pre-bounce phase of the bounce models, the universe
undergoes a contracting. It is shown that, in contrast with the
expanding phase, the vector mode will exhibit a growing behavior
\cite{r97}. The unlimited growth of the perturbation will breakdown
the perturbation theory. So, it is necessary to check the behaviors
of the vector mode near the bounce.

The effective linearized equations of vector mode with holonomy
corrections have been given in \cite{r96} (in Newton gauge):
\begin{widetext}
\begin{equation}
-\frac 1{2a^2}\nabla ^2{\cal S}^i=(8\pi G)^2(\rho +P){\cal V}^i,
\label{e99}
\end{equation}
\begin{equation}
-\frac 12\partial _\eta \left( {\cal S}_{~,j}^i+{\cal S}_{~,i}^j\right) -%
\frac{\bar{\mathfrak{K}}}{2a}\left( 1+\left[ \frac{\sin
(2\tilde{\mu}\gamma \bar{\mathfrak{K}})}{2\gamma
\tilde{\mu}\bar{\mathfrak{K}}}\right] \right)
\left( {\cal S}_{~,j}^i+{\cal S}_{~,i}^j\right) +{\cal{G}}%
_{(i}^{~~j)}=(8\pi G)^2\bar{p}\left( \Pi _{~,j}^i+\Pi
_{~,i}^j\right), \label{e98}
\end{equation}
\end{widetext}
where $\mathcal{S}^i$ and $\mathcal{V}^i$ are the metric
perturbation and the 4-velocity perturbation of the vector mode
respectively, $\Pi^i$ is the anisotropic stress, and
${\cal{G}}_{i}^{~j}$ is the anomaly term \cite{r96}. To have a
consistent set of the evolution equations, we require the anomaly
term to vanish i.e. ${\cal{G}}_{i}^{~j}=0$.

From Eq.(\ref{e99}), one can obtain a relation between $\mathcal{V}$
and $\mathcal{S}$ in Fourier mode of $k$:
\begin{equation}\label{e95}
\mathcal{V}_k^i=\frac{1}{2(8\pi G)^2 \bar{p}
(\rho+P)}k^2\mathcal{S}_k^i.
\end{equation}

If we do not take into account the anisotropic of perturbations, the
Eq.(\ref{e98}) will be:
\begin{equation}\label{e94}
\left[-\frac{1}{2}\partial_{\eta}-\frac{\bar{\mathfrak{K}}}
{2a}\left(1+\left[\frac{\sin
(2\tilde{\mu}\gamma\bar{\mathfrak{K}})}{2\gamma\tilde{\mu}\bar{\mathfrak{K}}}\right]
\right)\right]\left(\mathcal{S}^i_{~,j}+\mathcal{S}^j_{~,i}\right)=0,
\end{equation}
and the equation of the Fourier mode $k$ is
\begin{equation}\label{e93}
\frac{\partial}{\partial
\eta}\mathcal{S}^i_k+\frac{\bar{\mathfrak{K}}}{a}\left(1+
\left[\frac{\sin
(2\tilde{\mu}\gamma\bar{\mathfrak{K}})}{2\gamma\tilde{\mu}\bar{\mathfrak{K}}}\right]\right)\mathcal{S}^i_k=0.
\end{equation}
We will solve the Eq.(\ref{e93}) in two models which is the same as
the Sec. \ref{s3}.

\subsection{Massless scalar field}
For the same massless scalar field model discussed in Sec. \ref{s4},
the evolution of the scale factor is Eq.(\ref{e39}), and the
Eq.(\ref{e93}) changes into
\begin{equation}\label{e92}
\frac{\partial}{\partial
\eta}\mathcal{S}^i_k+\frac{\mathbb{A}^2}{\bar{p}_{min}}\eta\mathcal{S}^i_k=0.
\end{equation}
We only consider the leading term and thus
\begin{equation}\label{e91}
\mathcal{S}^i_k\propto \exp\left(-\frac{\mathbb{A}^2}{2\bar{p}_{min}}\eta^2\right)
\end{equation}
We can see that, even though the vector mode is growing when the
universe contracting to bounce, there is a maximum at the point of
the bounce. It means that the vector mode should not growing
unlimited.

As pointed out in \cite{r97} that, only the combination
$(\rho+P)\mathcal{V}^i$ appears in the energy momentum tensor;
therefore it is this combination that could in principle be
observable and may thus be called physically relevant.

From Eq.(\ref{e95}), we can obtain:
\begin{equation}\label{e90}
(\rho+P)\mathcal{V}^i_k\propto\frac{\exp\left(-\frac{\mathbb{A}^2}{2\bar{p}_{min}}\eta^2\right)}{(\bar{p}_{min}^2+\mathbb{A}\eta^2)^{1/2}}
\end{equation}
This also have a maximum at the point of bounce and it inversely
proportional with $\bar{p}_{min}=a_{bounce}^2$.

\subsection{Toy model}
For the same toy model discussed in Sec. \ref{s5}, the form of
$\bar{p}(\eta)$ is taken as Eq.(\ref{e49}) and the Eq.(\ref{e93})
approximating into the leading order becomes
\begin{equation}\label{e89}
\frac{\partial}{\partial\eta}\mathcal{S}^i_k+\frac{2\alpha}{b^m}\eta^{2m-1}\mathcal{S}^i_k=0.
\end{equation}
We can obtain
\begin{equation}\label{e88}
\mathcal{S}^i_k\propto
\exp\left(-\frac{\alpha}{mb^m}\eta^{2m}\right).
\end{equation}
This means that the limited growing is the same as Eq.(\ref{e91}),
and we also have
\begin{equation}\label{e87}
(\rho+P)\mathcal{V}^i_k\propto\frac{\exp\left(-\frac{\alpha}{mb^m}\eta^{2m}\right)}{(b^m+\alpha\eta^{2m})^{1/m}}.
\end{equation}

We find that, the maximum of $(\rho+P)\mathcal{V}^i_k$ is also
inversely proportional with $b=a_{bounce}^2$. It means that the
maximum of the observable quantity $(\rho+P)\mathcal{V}^i_k$ near
the bounce is inversely proportional to the square of scale factor
at the bounce point, and this conclusion is independent with the
model.

\section{\label{s7}discussion and conclusions}
In this paper, we examined the behaviors of the scalar and the
vector perturbations in the bounce phase of the effective theory of
LQC. Differing from the bounce model in \cite{r3}, the scalar
perturbations in our model is not divergence near both the bounce
point and the NEC transition point. Another conclusion is that the
vector mode of perturbations have maximum at the bounce point, and
this maximum is inversely proportional to the square scale factor at
the bounce point.

In the model of GR bounce, the emergence of bounce phase is rooted
in the matter in the universe. According to the singularity theorems
\cite{r2,r16}, if one requires the matter satisfies the energy
conditions, the universe can be emerged from an initial singularity.
However, there is no evidence that the exotic matter which violates
the NEC does not exist. So one can choose some exotic matter to make
the universe to experience a bounce. It is also because of this, the
behavior of the bounce and the perturbation near the bounce point is
decided by some exotic matter which have been chosen. Therefore, we
can select the matter carefully to make the behavior of perturbation
near the bounce have good performance like in \cite{r12,r13,r14}.
But too much artificial factors will make the physics of the model
unnatural.

On the other hand, the LQC bounce is originated in the discrete
spacetime geometry. Just like the model in Sec. \ref{s4}, even if
the matter satisfies the NEC, there is also a bounce phase. So the
behavior of the bounce is decided by the effects of discrete
spacetime geometry. From the analysis in Sec. \ref{s4} and \ref{s5},
one can find that, the effects of discrete spacetime geometry lead
to the convergence of the Bardeen potential.

One should note that, our discussion is in the framework of the
effective theory of LQC, so we find that this effective theory
reflects the nature of quantum spacetime geometry effectively.
Moreover, from the discussion of this paper, we also can obtain the
conclusion that the existence of LQC bounce is reasonable, it do not
lead to unbounded growth of the perturbation.

\acknowledgements

This work was supported by the National Natural Science Foundation
of China under Grant No. 10875012 and the Fundamental Research Funds
for the Central Universities.


\begin{thebibliography}{99}

\bibitem{r7}Borde A and Vilenkin A, Int. J. Mod. Phys. D {\bf 5} 813
(1996).

\bibitem{r8}Tolman R C, Phys. Rev.  {\bf 38} 1758
(1996).

\bibitem{r9}Durrer R and Laukenmann J, Class. Quantum Grav.  {\bf 13}
1069 (1996).

\bibitem{r10}Elbaz E, Novello M, Salim J M and Oliverira L A R, Int. J. Mod. Phys. D {\bf 1}
641 (1993).

\bibitem{r11}Patrick Peter and Nelson Pinto-Neto, Phys. Rev. D  {\bf 65}
023513 (2001).

\bibitem{r12}Patrick Peter, Nelson Pinto-Neto and Diego A Gonzalez, J. Cosmology and Astroparticle Phys.  {\bf 12}
003 (2003).

\bibitem{r13}Patrick Peter and Nelson Pinto-Neto, Phys. Rev. D  {\bf 66}
063509 (2002).

\bibitem{r14}Fabio Finelli, Patrick Peter and Nelson Pinto-Neto, Phys. Rev. D  {\bf 77}
103508 (2008).

\bibitem{r97}T. J. Battefeld and R. Brandenberger,
Phys. Rev. D {\bf 70}, 121302(R) (2004)

\bibitem{lqg1}T.Thiemann,\emph{Introduction to Modern Canaoical Quantum General Relativity} (Cambridge University Press, Cambridge, England,
2007).

\bibitem{lqg2}C.Rovell,\emph{Quantum Gravity} (Cambridge University Press, Cambridge, England,
2004).

\bibitem{lqg3}A.Ashtekar and J.Lewandowski, Class. Quantum Grav. {\bf 21},
R53 (2004).

\bibitem{B}M.Bojowald, Class. Quantum Grav. {\bf 17}, 1489 (2000); {\bf 17}, 1509 (2000); {\bf 18}, 1055
(2001); {\bf 18}, 1071 (2001).

\bibitem{bb}M.Bojowald, Phys. Rev. Lett {\bf 86},
5227-5230 (2001).

\bibitem{nbb1}A.Ashtekar, T.Pawlowski, P.Singh, Phys. Rev. D {\bf 73},
124038 (2006).

\bibitem{nbb2}A.Ashtekar, T.Pawlowski, P.Singh, Phys. Rev. D {\bf 74},
084003 (2006).

\bibitem{r15}Parampreet Singh, Kevin Vandersloot and G. V. Vereshchagin, Phys. Rev. D {\bf 74},
043510 (2006).

\bibitem{r99}J. Audretsch and G. Sch$\ddot{a}$fer, Phys. Lett. {\bf 66A} 459 (1978)

\bibitem{r98}N. Birrell and P. Davies, \emph{Quantum Fields in Curved Space}
(Cambridge University Press, Cambridge, England, 1982).

\bibitem{Li}Yu Li and Jian-Yang Zhu, Class. Quantum Grav. {\bf 28}, 045007 (2011)

\bibitem{r3}Patrick Peter, Nelson Pinto-Neto, Phys. Rev. D {\bf 65},
023513 (2001).

\bibitem{r4}S.Weinberg, \emph{Cosmology} (Oxford University Press, Oxford, England,
2008).

\bibitem{r1}Jian-Pin Wu and Yi Ling, J. Cosmology and Astroparticle
Phys. {\bf 05}, 026 (2010).

\bibitem{r6}J. Mielczarek, T. Stachowiak, and M.Szyd{\l}owski, Phys. Rev. D {\bf 77},
123506 (2008).

\bibitem{r96}M. Bojowald and G. M. Hossain, Class. Quantum  Grav. {\bf 24}, 4801 (2007)

\bibitem{r2}R.M.Wald, \emph{General Relativity} (Chicago University Press, Chicago,
1984).

\bibitem{r16}S. W. Hawking and G. F. R. Ellis, \emph{The Large Scale Structure of Space-time} (Cambridge Univerity Press, England,1973).


\end{thebibliography}
\end{document}